\documentclass[runningheads,a4paper]{llncs}
\usepackage{hyphenat}
\usepackage{color}
\usepackage{amssymb}
\usepackage{hyperref}
\hypersetup{
    colorlinks=true,
    linkcolor=black,
    filecolor=black,      
    urlcolor=blue,
    citecolor=black,
}
\usepackage{graphicx}
\usepackage[misc,geometry]{ifsym} 
\usepackage{amsmath}
\usepackage[colorinlistoftodos]{todonotes}

\usepackage{url}
\usepackage{hhline}
\usepackage{multirow}
\begin{document}

\mainmatter  

\title{Ultra2Speech - A Deep Learning Framework for Formant Frequency Estimation and Tracking from Ultrasound Tongue Images}
\titlerunning{Ultra2Speech}

\author{
Pramit Saha \inst{1}\textsuperscript{(\Letter)}
\and Yadong Liu \inst{2}
\and Bryan Gick \inst{2,3}
\and Sidney Fels \inst{1}
}

\authorrunning{P. Saha et al.}

\institute{Electrical and Computer Engineering Department, University of British Columbia\\
\and Department of Linguistics, University of British Columbia, Vancouver, Canada\\
\and Haskins Laboratories, New Haven, Connecticut, USA\\
\email{pramit@ece.ubc.ca}}


\toctitle{Ultra2Speech}

\tocauthor{**********, *********, *************}
\maketitle


\begin{abstract}
Thousands of individuals need surgical removal of their larynx due to critical diseases every year and therefore, require an alternative form of communication to articulate speech sounds after the loss of their voice box. This work addresses the articulatory-to-acoustic mapping problem based on ultrasound (US) tongue images for the development of a silent-speech interface (SSI) that can provide them with an assistance in their daily interactions. Our approach targets automatically extracting tongue movement information by selecting an optimal feature set from US images and mapping these features to the acoustic space.  We use a novel deep learning architecture to map US tongue images from the US probe placed beneath a subject's chin to formants that we call, Ultrasound2Formant (U2F) Net. It uses hybrid spatio-temporal 3D convolutions followed by feature shuffling, for the estimation and tracking of vowel formants from US images. The formant values are then utilized to synthesize continuous time-varying vowel trajectories, via Klatt Synthesizer. Our best model achieves R-squared ($R^2$) measure of 99.96\% for the regression task. Our network lays the foundation for an SSI as it successfully tracks the tongue contour automatically as an internal representation without any explicit annotation. 

\end{abstract}


\section{Introduction}

Human speech is a spontaneous yet powerful mode of communication. But millions of people fail to communicate through vocalization, due to severe diseases and speech disorders. One of the key components of speech production is the vocal fold, housed within larynx, which is responsible for providing the major source excitation for speech through its vibrations. Many people need to undergo laryngectomy or surgical removal of larynx for treating laryngeal cancer, critical neck injuries and radio-necrosis of the larynx. Though laryngeal cancer accounts for only 1\% of all cancers, it has around 70\% 5-year survival rate \cite{gilbert2017restoring}. Those undergoing laryngectomy are limited to speak sub-vocally by moving their vocal tract articulators including their tongue, without engaging their vocal fold. However, the loss of voice and failure to effectively communicate owing to the lack of a voice-box, can have a devastating impact in the quality of life of post-laryngectomy patients. The commercially available technologies for voice rehabilitation, including trachea-esophageal speech and electrolarynx speech, have significant limitations as discussed in \cite{gilbert2017restoring}. Therefore, there is a need for an alternative form of personalized communication device for such patients, that do not rely on the acoustic signals to produce speech. The ability to communicate in the absence of acoustic signals can be facilitated by sensing the movement of the remaining speech articulators and converting those to speech. Such devices, also known as “silent speech interfaces" (SSI) \cite{denby2010silent}, involve the extraction of speech information via electro-encephalography, surface electromyography, ultrasound imaging, electromagnetic articulography, \textit{etc}. There are two distinct approaches of providing speech outputs from these interfaces, \textit{viz.} `recognition' and `direct synthesis'. Despite the rapid progress in speech token `recognition' led by the adoption of deep learning based classification \cite{saha2018towards}, attempts in speech `synthesis' has been less frequent due to the lack of efficient speech generation techniques.

In this paper, we introduce a novel ultrasound (US) based sound synthesis approach via a 3D convolutional neural network and formant based speech synthesis engine.
The proposed Ultra2Formant Net (U2F) presents a hybrid 3D convolutional block which involves the parallel decomposition of a chunk of standard 3D convolution into individual 2D spatial and 1D temporal convolutional filters. This constrained approach takes advantage of the orthogonal nature of spatial and temporal kernels to decrease the parameter set as well as increase the strength of net spatio-temporal feature encodings. These encodings are then combined with parallel 3D convolutional output, followed by rigorous feature shuffling. The network outputs the formant frequencies, which are then fed to the Klatt speech synthesis engine for generating desired continuous speech sounds. Our codes and other details are made available at \href{https://pramitsaha.github.io}{\url{https://pramitsaha.github.io/}}.

\section{Background and related works}

\subsection{Previous works}

The research in this field was initiated with an attempt \cite{denby2004speech} to synthesize speech based on 12 GSM vocoder parameters from tongue contour points using Multilayer-perceptrons (MLP). The input was later modified and combined with lip coordinates and mapped to 12 line spectral frequencies (LSF) using similar MLP networks \cite{denby2006prospects}. The input space was further altered in \cite{hueber2007eigentongue} to additionally include \textit{Eigentongue} features as well. Another related work \cite{csapo2017dnn} utilized different combinations of feature representation including eigen-tongue and correlation-based features to map to 13 MGC-LSP features using 5 layered DNN. In a follow-up work \cite{gosztolya2019autoencoder}, the authors replaced the hand-engineered features by an autoencoder, whose bottleneck features were then fed to the DNN layers for mapping to the MGC-LSP. In order to take advantage of a shared representation, a multi-task DNN was further employed in \cite{toth2018multi} to simultaneously classify phone states and synthesize spectral parameters. The latest work in this direction has been the application of CNN-LSTMs on US images denoised via convolutional autoencoders, intended to predict 24 order MGC-LSP coefficients for synthesizing speech in Hungarian language \cite{juanpere2019ultrasound}. However, the vowel transitions in the synthesized speech differ considerably from the desired speech, leading to the poor performance of the synthesized version. A promising way of achieving closer vowel trajectories is to explore the formant frequency space. The connection of articulatory space with the 2D formant space is particularly important to explore because the formant representation is a very powerful acoustic encoding that efficiently describes the essential aspects of speech using limited parameters. It also provides a lot of insight on continuous vowel trajectories, the most dynamic part of speech production, that can be utilized for better control of speech in SSI. To the best of our knowledge, this is the first investigation on deep learning based ultrasound-to-formant mapping for speech synthesis.
\subsection{Formant estimation and tracking}
Estimation and tracking of formant frequency is one of the fundamental problems in speech processing \cite{oshaughnessy2008formant}. This involves determining the formant frequencies corresponding to stationary speech segment and tracking these throughout the signal. Since these formant frequencies are the direct results of resonances brought about by the tongue movement, the problem somewhat boils down to identifying and tracking the tongue contour. This is, to some extent, analogous to the applications like video object tracking, action recognition, \textit{etc.} which utilize different spatio-temporal feature encoding schemes \cite{tran2018closer,carreira2017quo,mandal2020motionrec,luo2019grouped}. However, there are numerous other challenges encountered in ultrasound based tongue localizing and tracking. For example, we cannot use pre-trained networks popular in video processing as backbones for our application. Besides, the ultrasound images are grayscale, contain less information, possess low spatial resolution and are infested with noises and artifacts.

\subsection{Challenges in tongue tracking}

Tongue is a muscular hydrostat having no conventional skeletal support, which results in its remarkably diverse and complex movements \cite{stavness2012automatic}. 
Having multiple degrees of freedom, different parts of the tongue can move simultaneously towards different directions. As such, each tiny movement or shape change of the tongue results in corresponding changes of the vocal tract resonances, which in turn, changes the formant values at that time instant. However, there is a dearth of tongue contour annotations and lack of fully-automated, generalizable contour extraction methods that makes the U2S task much more challenging. As a result, it is crucial for a successful U2S mapping algorithm to be able to automatically track the tongue contour as a hidden representation, in order to understand the variation of formants from ultrasound.

\begin{figure}[t]

\includegraphics[height=.43\textwidth]{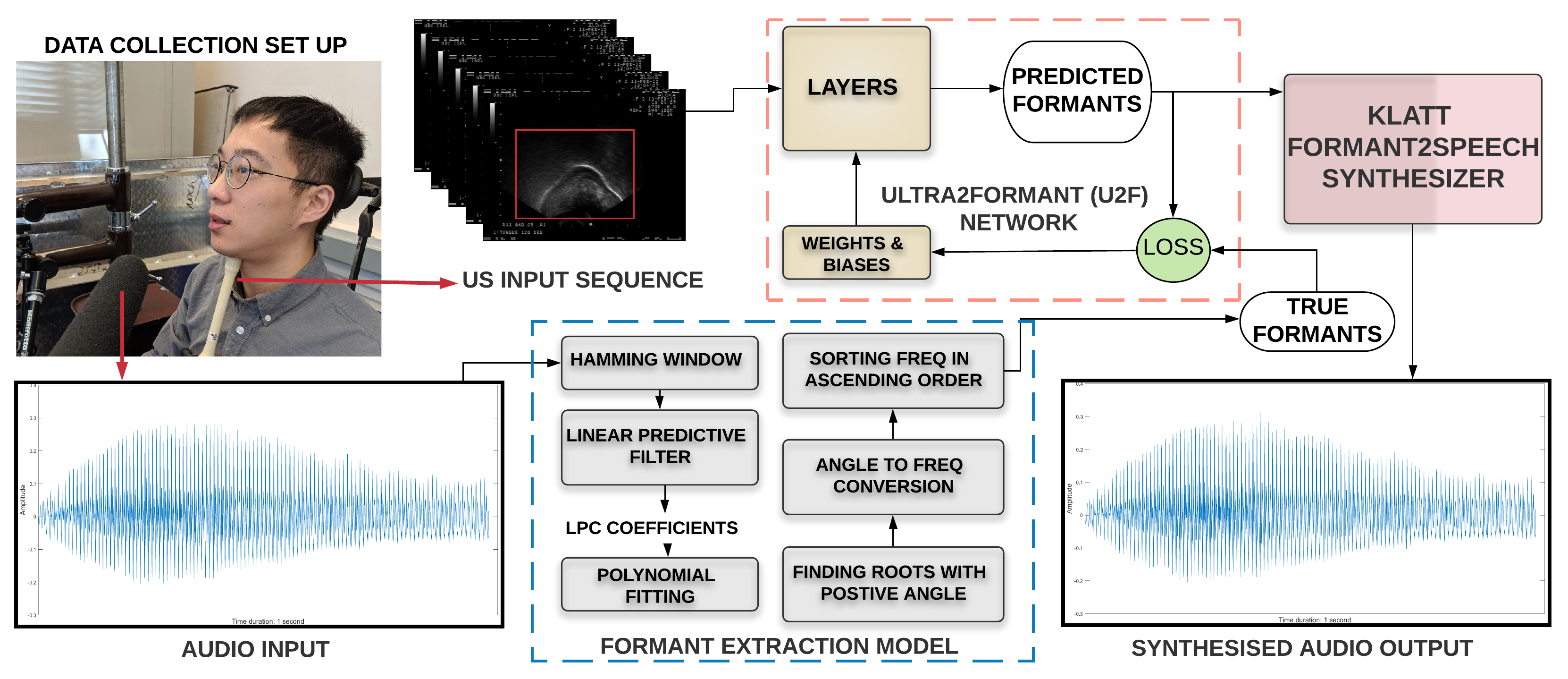}
\caption{Overview of proposed Ultra2Speech. The arrows indicate the data flow.}

\end{figure}
\section{Proposed Ultra2Speech (U2S) model}
We face three fundamental challenges in US-based formant estimation and tracking: (1) extracting relevant spatial information for accurate tongue contour detection; (2) encoding temporal information for understanding the dynamics of tongue movement; and (3) reaching the desired mapping between the extracted spatio-temporal features and the formant trajectories. In this section, we first introduce our \textbf{Ultra2Formant (U2F) Net}, aimed at tackling these challenges using different kernels of 3D CNNs, as illustrated in Fig 2.

%
\subsection{First hidden layer} 
The input video is first convolved with a set of pointwise convolutional filters with kernel of size $1\times 1 \times 1$.  Such filters are known to reduce the computational complexity before expensive $3\times 3 \times 3$ operations. Besides, they also extract efficient low dimensional embedding and apply extra non-linear activations that help the network to model complex functions.

\subsection{Hybrid convolutional layer}
The output channels of pointwise convolutions are split into three groups, one for intra-frame spatial feature extraction, another for cross-frame temporal modeling and the other for joint spatio-temporal encoding. The spatial branch is composed of 2D CNN kernels $1\times3\times3$; the temporal branch is composed of 1D CNN time-kernels $3\times1\times1$; and the joint spatio-temporal branch is composed of 3D CNN kernels $3\times3\times3$. In this way, we constrain some particular feature channels to focus more on static spatial features, while few others to focus on dynamic motion representation and remaining on encoding joint information. Factorizing part of the standard 3D convolution kernel\cite{luo2019grouped} into orthogonal parallel components reduces the number of parameters thereby making it easier to train. Besides, the separation of orthogonal features also contributes towards better optimization of loss function, as reflected in the performance later. This partial decoupling of the spatial and temporal kernels of 3D CNN makes it both effective in performance and efficient in computation. 


\subsection{Feature shuffling, grouped convolution and fully connected layer}
The output from three branches are concatenated together and shuffled in three groups. Consider the concatenated features with 3 groups, each having N channels. For shuffling, the output channel dimension is first reshaped into (3, N),  followed by transposing and then flattening it back to its previous shape for feeding it to the next layer. This shuffling facilitates cross-group information exchange and strengthens the spatio-temporal encoding within a computational budget as shown in a different context in \cite{zhang2018shufflenet}. The feature representation is further compressed by passing it through a grouped convolutional layer of kernel size $1 \times 1\times 1$. The output features are finally flattened and connected to two parallel sets of 30 output nodes through task-specific fully-connected layers. This joint learning paradigm aimed at estimating two sets of formant frequencies using the same network creates a shared representation beneficial for the model. 

\begin{figure}[t]
\centering
\includegraphics[height=.55\textwidth]{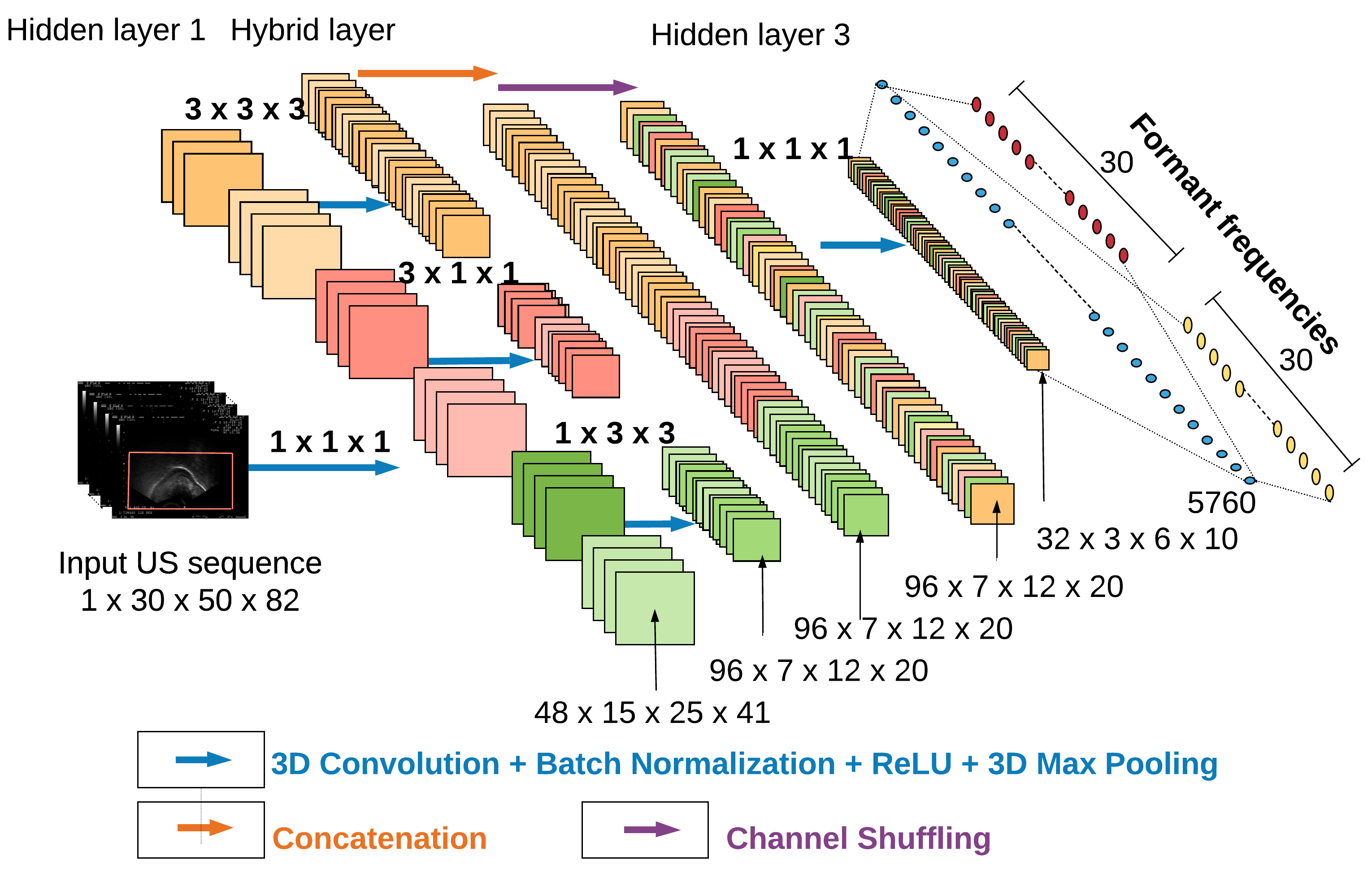}
\caption{Architecture of the proposed Ultra2Formant (U2F) Net}

\end{figure}


\subsection{Formant2speech block}
We utilize Klatt synthesis software that accepts the U2F outputs (formants) as its input parameters and generates speech output as shown in Fig 1. Due to space constraints, we refer the readers to \cite{klatt1980software} for further study on the joint parallel-cascaded formant-to-speech synthesis methodology. 
\section{Experiment and results}
\subsection{Data acquisition and preprocessing}
 
\subsubsection{Ultrasound and audio data collection:}

Since our target was to develop a personalized SSI, we collected mid-sagittal US videos of a single male participant over a number of sessions.
Throughout the data collection procedure, the participant was seated with his head stabilized against a headset and was asked to make continuous open vocal tract sounds with intervals in between. For imaging the tongue, the ultrasound transducer was placed beneath the chin. 
The imaging was done using ALOKA SSD-5000 ultrasound system at 30 fps, with 180 degree 9mm radius UST-9118 3.5 MHz convex ultrasound probe. 
Mono-channel audio recording was done simultaneously using Praat software, a Sennheiser MKH 416 P48 shotgun microphone and a Focusrite Scalet 2i2 preamplifier. In order to align the audio soundtrack and the video recording, participants were asked to produce sounds that involves a sudden salient acoustic change and a quick noticeable tongue movement, such as \textit{/ga/}.
\subsubsection{Audio-visual alignment, image extraction and pre-processing:}
Since the audio recording started prior to video recording, there was a time lag between audio and video recordings which ought to be calculated. For this, the frame of production of the release of \textit{/ga/} in ultrasound imaging data and the timestamp of the same event in the audio recording were identified and used for synchronization. 
Vowel sequences were identified and segmented from the audio recordings, and acoustic landmarks were prepared. Further, ultrasound video recording were converted to image sequences at 30 fps using QuickTime 7 Pro and the frames corresponding to each vowel sequence were extracted considering the time of acoustic landmarks and the lag between audio and video recordings. 
The frames of spatial resolution $ 480 \times 640 $ were cropped using a bounding box of $200 \times 330$ that contained the tongue for the entire image sequence and were further downsampled into  $ 50 \times 82 $. We also converted the images to grayscale and normalized the pixel intensities within $[0,1]$. We chose a time window of 30 frames, resulting in a total of 13,082 videos of duration 1 second each. This time window was chosen as a trade-off between the dynamic information to be imparted to the network and the computational time required for training.

\subsubsection{Formant extraction from recorded speech:} We applied a Hamming window on frame-blocked acoustic signal of 1470 samples each. Following the traditional approach, next, we employed a 1D filter with transfer function of $1/(1+0.63z^-1)$ and computed the linear predictive coefficients for the filtered signal using autocorrelation method. Furthermore, we computed the roots of the predictor polynomial in order to locate the peaks in spectra of LPC filters. Only positive frequencies up to half of the sampling frequency were considered for calculations and were sorted in ascending order. In this study, we explored the first two formant frequencies - the most dominating parameters for speech trajectories.  

\subsection{Implementation details and performance analysis}
\subsubsection{Network architecture:}
The final model consists of 4 hidden layers, 3 being convolutional, with respectively 48, 96 and 32 filters and the last one being a fully-connected layer of 5760 nodes, connected parallely to the output nodes as illustrated in Fig 2. Each 3D convolutional layer has a stride=1 and is followed by 3D Batch Normalization\cite{ioffe2015batch}, ReLU activation\cite{dahl2013improving} and 3D max-pooling with a stride=2. We perform a respective padding of (0,1,1), (1,0,0) and (1,1,1) for the spatial, temporal and spatio-temporal convolution of the hybrid layer. 


\subsubsection{Training  and evaluation:} Our U2S model was implemented in PyTorch. We randomly shuffled and partitioned the data (13,082 videos) into train (80\%), development (10\%) and test sets (10\%). The network was trained with a batch size of 10 on NVIDIA GeForce GTX 1080 Ti GPU. Mean absolute error (MAE) loss function was optimized using Adam with a learning rate of .001 for a total of 100 epochs. All the parameters were randomly initialized. In order to mitigate the problem of overfitting, we used Batch Normalization after every convolutional layer and before applying non-linearity. The architectural parameters and hyperparameters shown in Fig 2 were selected through an exhaustive grid-search. Since the results were computed as a sequence of f1 and f2 values (30 each) as shown in Fig 3 (a), we used Mean Absolute Error (MAE) and Mean R-squared ($R^2$) as metrics for quantifying the regression performance.

\begin{figure}[t]
\centering
\includegraphics[height=.22\textwidth]{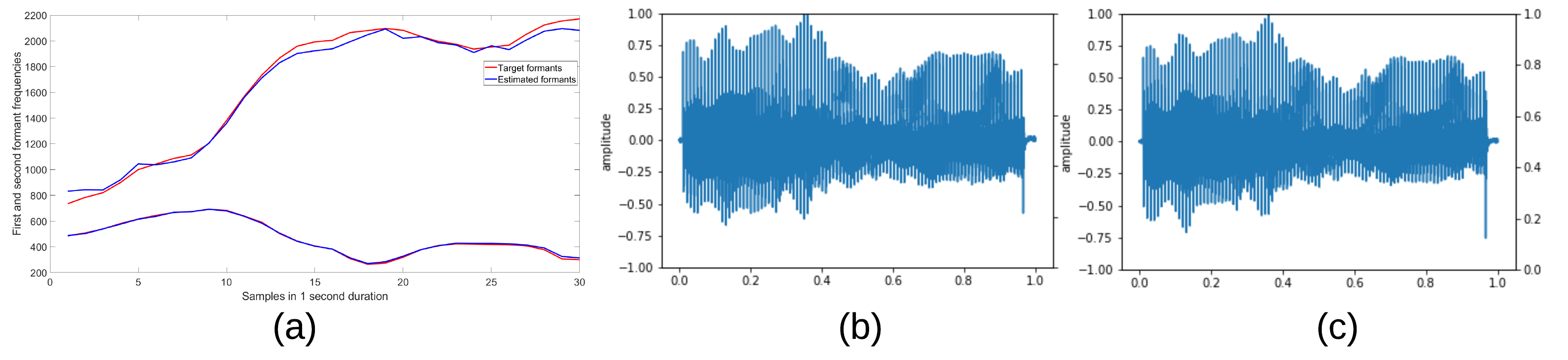}
\caption{(a) Time-varying formants (Red indicates target and blue indicates predicted trajectories), (b) Original speech signal, (c) synthesized speech signal}

\end{figure}
\begin{figure}[t]
\centering
\includegraphics[height=.12\textwidth]{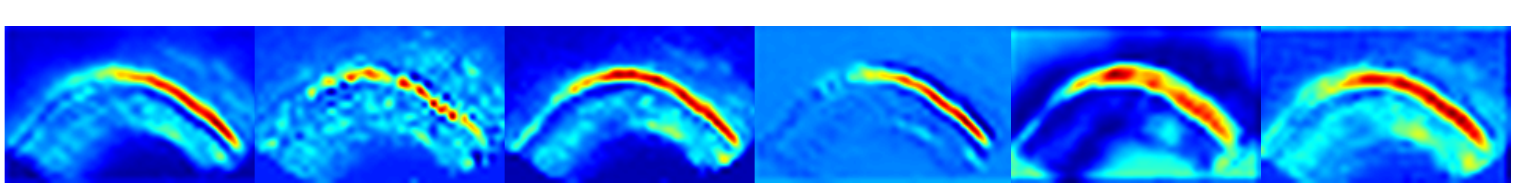}
\caption{Saliency maps from U2F showing internal tongue contour localization}

\end{figure}

\subsubsection{Results:} We showcase our results on a randomly chosen sample in Fig 3, which demonstrates that there is almost no visible distinction between the target and predicted acoustic signal. We also show the visual explanation behind the estimations made by U2F in Fig 4 in the form of saliency maps corresponding to last CNN layer. This surprisingly reveals its striking ability to accurately represent the tongue contour internally.   
Table 1 presents the quantitative results and comparisons, corresponding to a joint f1-f2 prediction task Vs individual formant prediction task. The joint configuration consistently achieves better performance taking advantage of a shared representation, despite having fewer parameters. Our baseline method is the Conv-BiLSTM that has been presented as the state-of-the-art approach in \cite{juanpere2019ultrasound}. The network in the fourth row of the Table 1  has exactly same architecture as U2F except that the hybrid CNN-block is replaced by regular CNN-block. The results show that the proposed U2F model outperforms the CNN-RNN baselines as well as the standard 3D CNN models even with lesser number of parameters. 
\begin{table}
  \centering
  \caption{Performance comparison with baseline methods}
  \renewcommand{\arraystretch}{1}
  \begin{tabular}{|p{4.2cm}|c|c|c|c|c|c|}
    \hline
    \multirow{1}{2cm}{\textbf{Method}} & \multicolumn{2}{c|}{\textbf{$f1$}} & \multicolumn{2}{c|} {\textbf{$f2$}}&  \multicolumn{2}{c|}{\textbf{$f1-f2$}}\\
    \cline{2-7}
    & \textbf{MAE} & \textbf{ $Mean~R^2$} & \textbf{MAE} & \textbf{$Mean~ R^2$}& \textbf{MAE} & \textbf{$Mean~R^2$}\\

    \hline\hline
    CLSTM(2-layers)+2-FCN & .0419 & 86.36 & .0423 & 85.34 & .0444 & 86.45 \\ \hline
    CBiLSTM(2-layers)+2-FCN &.0352 & 89.40 & .0380 & 89.12 & .0293  & 90.01 \\ \hline 
    3D CNN(2-layers)+1-FCN & .0233 & 96.79 & .0242 & 96.22 & .0069 & 98.87  \\ \hline
    3D CNN(4-layers)+1-FCN & .0204 & 98.40 & .0174 & 98.13 & .0118 & 98.78  \\ \hline
    Our U2F (with Hybrid Conv) & \textbf{.0097} & \textbf{99.80} & \textbf{.0092} & \textbf{99.76} & \textbf{.0052} & \textbf{99.96} \\ \hline
  \end{tabular}
  
\end{table}

\subsubsection{Ablation study}
We conduct several ablation experiments on our dataset to analyze the contribution of different modules of U2F Net. 
Here, we particularly report  (in Table 2) three primary variants of our model obtained by dropping the spatial layer, the temporal layer, and the channel shuffling. We can see that the network performance decreases by 1.88\% and 1.44\% in absence of the individual spatial and temporal encoding respectively. This shows that the hybrid block is a significant part of U2F which captures contrasting features to jointly learn the localization and tracking of tongue contour better. 
Similarly, the removal of shuffling block leads to an approximate decrease of the Mean $R^2$ by .84\%. This is because channel shuffling mixes the independent as well as shared encodings and thereby enriches the input feature space for the last grouped CNN layer.
All the ablation studies provide evidence in favour of our original model design. 

\begin{table}
   \centering
  \caption{Ablation experiments - \textit{Removal of spatial, temporal and shuffling blocks}}
  \renewcommand{\arraystretch}{1}
  \begin{tabular}{|p{4cm}|c|c|c|c|c|c|}
    \hline
    \multirow{1}{2cm}{\textbf{Method}} & \multicolumn{2}{c|}{\textbf{$f1$}} & \multicolumn{2}{c|} {\textbf{$f2$}}&  \multicolumn{2}{c|}{\textbf{$f1-f2$}}\\
    \cline{2-7}
    & \textbf{MAE} & \textbf{ $Mean~R^2$} & \textbf{MAE} & \textbf{$Mean~ R^2$}& \textbf{MAE} & \textbf{$Mean~R^2$}\\

    \hline\hline
    U2F w/o spatial kernels & .0178 & 97.78 & .0267 & 97.69 & .0113 & 98.08 \\ \hline
    U2F w/o temporal kernels & .0188 & 98.01 & .0180 & 98.30 & .0161 & 98.52 \\ \hline
    U2F w/o shuffling block  & .0103 & 98.84 & .0097 & 99.09 & .0087 & 99.12  \\ \hline

  \end{tabular}
  
\end{table}


 \section{Discussion and Conclusion}
  The main contributions of our paper are four-fold: 
 \begin{enumerate}

 \item We developed a novel spatio-temporal feature extraction strategy for mapping ultrasound tongue movement to formant trajectories. This involves replacing a chunk of the 3D convolutional layer by individual 2D spatial and 1D temporal convolutions for better feature encoding. A shuffling block is introduced to enable cross-feature information flow between spatial, temporal and spatio-temporal representations. 
 
 \item For the first time, we established a successful end-to-end mapping between the ultrasound tongue images and formant frequencies, that bridges the gap in SSI and opens a new dimension for articulatory speech research. 
 
\item We provide evidences that our network has the ability to model an internal representation of tongue by optimizing a non-image based loss function. This demonstrates that the network has the potential to replace the manual selection of points for semi-automatic tongue contour extraction. This also shows the promise of using acoustic labels for tongue contour detection, thereby, replacing the need for tedious manual annotation for tongue tracing. 

\item Our approach shows a striking improvement in performance over the baseline methods. We present an ablation study to explain the contribution of individual components towards better performance. Our network has the potential to encode robust spatio-temporal information in other related tasks.
 \end{enumerate}
 
\noindent\textbf{Acknowledgements.}
This work was funded by the Natural Sciences and Engineering Research Council (NSERC) of Canada and Canadian Institutes for Health Research (CIHR).
\bibliography{myBib} {}

\begin{thebibliography}{10}
\providecommand{\url}[1]{\texttt{#1}}
\providecommand{\urlprefix}{URL }
\providecommand{\doi}[1]{https://doi.org/#1}

\bibitem{carreira2017quo}
Carreira, J., Zisserman, A.: Quo vadis, action recognition? a new model and the
  kinetics dataset. In: proceedings of the IEEE Conference on Computer Vision
  and Pattern Recognition. pp. 6299--6308 (2017)

\bibitem{csapo2017dnn}
Csap{\'o}, T.G., Gr{\'o}sz, T., Gosztolya, G., T{\'o}th, L., Mark{\'o}, A.:
  Dnn-based ultrasound-to-speech conversion for a silent speech interface.
  Proc. Interspeech 2017 pp. 3672--3676 (2017)

\bibitem{dahl2013improving}
Dahl, G.E., Sainath, T.N., Hinton, G.E.: Improving deep neural networks for
  lvcsr using rectified linear units and dropout. In: 2013 IEEE international
  conference on acoustics, speech and signal processing. pp. 8609--8613. IEEE
  (2013)

\bibitem{denby2006prospects}
Denby, B., Oussar, Y., Dreyfus, G., Stone, M.: Prospects for a silent speech
  interface using ultrasound imaging. In: 2006 IEEE International Conference on
  Acoustics Speech and Signal Processing Proceedings. vol.~1, pp.~I--I. IEEE
  (2006)

\bibitem{denby2010silent}
Denby, B., Schultz, T., Honda, K., Hueber, T., Gilbert, J.M., Brumberg, J.S.:
  Silent speech interfaces. Speech Communication  \textbf{52}(4),  270--287
  (2010)

\bibitem{denby2004speech}
Denby, B., Stone, M.: Speech synthesis from real time ultrasound images of the
  tongue. In: 2004 IEEE International Conference on Acoustics, Speech, and
  Signal Processing. vol.~1, pp. I--685. IEEE (2004)

\bibitem{gilbert2017restoring}
Gilbert, J.M., Gonzalez, J.A., Cheah, L.A., Ell, S.R., Green, P., Moore, R.K.,
  Holdsworth, E.: Restoring speech following total removal of the larynx by a
  learned transformation from sensor data to acoustics. The Journal of the
  Acoustical Society of America  \textbf{141}(3),  EL307--EL313 (2017)

\bibitem{gosztolya2019autoencoder}
Gosztolya, G., Pint{\'e}r, {\'A}., T{\'o}th, L., Gr{\'o}sz, T., Mark{\'o}, A.,
  Csap{\'o}, T.G.: Autoencoder-based articulatory-to-acoustic mapping for
  ultrasound silent speech interfaces. In: 2019 International Joint Conference
  on Neural Networks (IJCNN). pp.~1--8. IEEE (2019)

\bibitem{hueber2007eigentongue}
Hueber, T., Aversano, G., Cholle, G., Denby, B., Dreyfus, G., Oussar, Y.,
  Roussel, P., Stone, M.: Eigentongue feature extraction for an
  ultrasound-based silent speech interface. In: 2007 IEEE International
  Conference on Acoustics, Speech and Signal Processing-ICASSP'07. vol.~1, pp.
  I--1245. IEEE (2007)

\bibitem{ioffe2015batch}
Ioffe, S., Szegedy, C.: Batch normalization: Accelerating deep network training
  by reducing internal covariate shift. arXiv preprint arXiv:1502.03167  (2015)

\bibitem{juanpere2019ultrasound}
Juanpere, E.M., Csap{\'o}, T.G.: Ultrasound-based silent speech interface using
  convolutional and recurrent neural networks. Acta Acustica united with
  Acustica  \textbf{105}(4),  587--590 (2019)

\bibitem{klatt1980software}
Klatt, D.H.: Software for a cascade/parallel formant synthesizer. the Journal
  of the Acoustical Society of America  \textbf{67}(3),  971--995 (1980)

\bibitem{luo2019grouped}
Luo, C., Yuille, A.L.: Grouped spatial-temporal aggregation for efficient
  action recognition. In: Proceedings of the IEEE International Conference on
  Computer Vision. pp. 5512--5521 (2019)

\bibitem{mandal2020motionrec}
Mandal, M., Kumar, L.K., Saran, M.S., et~al.: Motionrec: A unified deep
  framework for moving object recognition. In: The IEEE Winter Conference on
  Applications of Computer Vision. pp. 2734--2743 (2020)

\bibitem{oshaughnessy2008formant}
OShaughnessy, D.: Formant estimation and tracking. In: Springer handbook of
  speech processing, pp. 213--228. Springer (2008)

\bibitem{saha2018towards}
Saha, P., Srungarapu, P., Fels, S.: Towards automatic speech identification
  from vocal tract shape dynamics in real-time mri. Proc. Interspeech 2018 pp.
  1249--1253 (2018)

\bibitem{stavness2012automatic}
Stavness, I., Lloyd, J.E., Fels, S.: Automatic prediction of tongue muscle
  activations using a finite element model. Journal of biomechanics
  \textbf{45}(16),  2841--2848 (2012)

\bibitem{toth2018multi}
T{\'o}th, L., Gosztolya, G., Gr{\'o}sz, T., Mark{\'o}, A., Csap{\'o}, T.G.:
  Multi-task learning of speech recognition and speech synthesis parameters for
  ultrasound-based silent speech interfaces. In: Interspeech. pp. 3172--3176
  (2018)

\bibitem{tran2018closer}
Tran, D., Wang, H., Torresani, L., Ray, J., LeCun, Y., Paluri, M.: A closer
  look at spatiotemporal convolutions for action recognition. In: Proceedings
  of the IEEE conference on Computer Vision and Pattern Recognition. pp.
  6450--6459 (2018)

\bibitem{zhang2018shufflenet}
Zhang, X., Zhou, X., Lin, M., Sun, J.: Shufflenet: An extremely efficient
  convolutional neural network for mobile devices. In: Proceedings of the IEEE
  conference on computer vision and pattern recognition. pp. 6848--6856 (2018)

\end{thebibliography}
\bibliographystyle{splncs04}

\end{document}